\documentclass[%
 aip,
 amsmath,amssymb,
 reprint,%
]{revtex4-1}

\usepackage{graphicx}% Include figure files
\usepackage{dcolumn}% Align table columns on decimal point
\usepackage{bm}% bold math
\usepackage[utf8]{inputenc}
\usepackage[T1]{fontenc}
\usepackage{mathptmx}
\usepackage{etoolbox}

\makeatletter
\def\@email#1#2{%
 \endgroup
 \patchcmd{\titleblock@produce}
  {\frontmatter@RRAPformat}
  {\frontmatter@RRAPformat{\produce@RRAP{*#1\href{mailto:#2}{#2}}}\frontmatter@RRAPformat}
  {}{}
}%
\makeatother
\begin{document}

% \preprint{AIP/123-QED}

\title[Node pruning reveals substructures]{Node pruning reveals compact and optimal substructures within large networks}

\author{Manish Yadav}%
\homepage{Corresponding Author}
\email{manish.yadav@tu-berlin.de}
\affiliation{Chair of Cyber-Physical Systems in Mechanical Engineering, Technische Universität Berlin, Straße des 17. Juni 135, 10623 Berlin, Germany}%
 
\author{Merten Stender}
\email{merten.stender@tu-berlin.de}
\affiliation{Chair of Cyber-Physical Systems in Mechanical Engineering, Technische Universität Berlin, Straße des 17. Juni 135, 10623 Berlin, Germany}%
% Force line breaks with \\
% \author{A. Author}
%  \altaffiliation[Also at ]{Physics Department, XYZ University.}%Lines break automatically or can be forced with \\
% \author{B. Author}%
%  \email{Second.Author@institution.edu.}
% \affiliation{ 
% Authors' institution and/or address%\\This line break forced with \textbackslash\textbackslash
% }%

% \author{C. Author}
%  \homepage{http://www.Second.institution.edu/~Charlie.Author.}
% \affiliation{%
% Second institution and/or address%\\This line break forced% with \\
% }%

\date{\today}% It is always \today, today,
             %  but any date may be explicitly specified

\begin{abstract}
The structural complexity of reservoir networks poses a significant challenge, often leading to excessive computational costs and suboptimal performance. In this study, we introduce a systematic, task-specific node pruning framework that enhances both the efficiency and adaptability of reservoir networks. By identifying and eliminating redundant nodes, we demonstrate that large networks can be compressed while preserving—or even improving—performance on key computational tasks. Our findings reveal the emergence of optimal subnetwork structures from larger Erdős-Rényi random networks, indicating that efficiency is governed not merely by size but by topological organization. A detailed analysis of network structure at both global and node levels uncovers the role of density distributions, special-radius and asymmetric input-output node distributions, among other graph-theoretic measures that enhance the computational capacity of pruned compact networks. We show that pruning leads to non-uniform network refinements, where specific nodes and connectivity patterns become critical for information flow and memory retention. This work offers fundamental insights into how structural optimization influences reservoir dynamics, providing a pathway toward designing more efficient, scalable, and interpretable machine learning architectures.
\end{abstract}

\maketitle

\begin{quotation}
Complex networks are fundamental to understanding interactions in diverse domains, from biological systems to technological infrastructures. Identifying compact and optimal substructures within these networks remains a key challenge in network science. In this study, we introduce a node pruning approach to systematically reveal critical substructures while preserving essential network properties. By applying our method to large Erdős-Rényi random networks, we demonstrate its effectiveness in uncovering underlying functional motifs and enhancing computational efficiency. These findings provide new insights into the structural organization of complex systems and offer practical implications for optimizing network-based processes.
\end{quotation}

\section{\label{sec:level1}Introduction}

In natural and artificial systems, network efficiency often emerges not from maximizing size but from optimizing structure. Biological networks, from neural circuits to metabolic pathways, demonstrate that excessive connectivity can lead to inefficiencies, such as increased energy consumption, slower response times, and reduced adaptability. For instance, in brain development, synaptic overproduction is followed by extensive pruning, refining functional connectivity while improving cognitive efficiency \cite{Selemon2013}. Similarly, metabolic networks in organisms optimize resource allocation by selectively eliminating redundant reactions, leading to more efficient biochemical pathways \cite{Barabasi2004}. Even at the genetic level, regulatory networks undergo structural refinement to enhance stability and adaptability, with excessive complexity often linked to dysfunction rather than improved function \cite{Alon2007}. Additionally, modular brain networks balance integration and segregation, allowing efficient information processing by reducing redundant connections while preserving critical functional pathways \cite{Sporns2016}.

How can one determine the optimal network size and structure to achieve a specific functionality, such as solving a predefined task? One approach is performance-dependent growth, where a small, minimally connected network is selectively expanded while improving task performance. This strategy, widely observed in biological development, has also been applied in artificial systems. For example, in neural network training, structured growth mechanisms allow networks to develop efficient architectures without unnecessary redundancy \cite{Bellec2018}. In reservoir computing, Yadav et al. (2025) demonstrated that recurrent networks can evolve through performance-driven expansion, selectively adding nodes and connections that improve task-specific processing while maintaining computational efficiency \cite{EvolveRC_Yadav2025}. A similar strategy explores how artificial neural networks can evolve from small initial configurations toward minimal optimal structures that maximize efficiency for a given task \cite{Sinha2025}. These studies showcase that the formation of smaller yet efficient is possible by performance-dependently increasing the network size.\\

However, an equally important yet underexplored strategy is performance-dependent pruning—starting with a large, randomly connected network and systematically reducing its size while preserving or even enhancing functionality. Unlike growth-based approaches, which build complexity from simplicity, pruning focuses on eliminating unnecessary components from an initially overconnected system. This principle is evident across various domains: in neuroscience, synaptic pruning refines neural circuits to improve efficiency \cite{Selemon2013}; in systems biology, network sparsification enhances metabolic and regulatory optimization \cite{Barabasi2004}; and in physics, sparsification techniques identify dominant interaction structures while reducing computational overhead \cite{Rocks2019, Spielman2011}.\\

Artificial systems also benefit from pruning-based optimization. In machine learning, structured pruning has shown that deep neural networks can significantly reduce size without compromising performance—and in some cases, even improving generalization by reducing overfitting \cite{Han2015, Molchanov2017}. Similarly, integrated circuits are designed by removing excess connections to improve signal transmission and reduce power consumption \cite{Mead1990}.
Pruning techniques have been widely explored in machine learning (ML) and artificial intelligence (AI) to reduce model complexity, improve efficiency, and maintain predictive performance. Early pruning methods were introduced in feedforward and convolutional neural networks (CNNs) to remove redundant weights and neurons. More recent work extends pruning strategies to recurrent networks, including Echo State Networks (ESNs) and other forms of Reservoir Computing (RC).  

Optimization-based approaches \cite{Hussain2023} focus on obtaining sparse yet effective networks by iteratively removing less critical connections. Complementary to this, perturbation-based pruning \cite{Yan2023} identifies unnecessary nodes through controlled disturbances in the network, revealing structurally weak or redundant components. Both approaches aim to retain model accuracy while reducing computational cost. One challenge in pruning is maintaining a degree of adaptability in the network after reduction. Recent studies \cite{Jiang2024} explore methods that allow the network to retain some plasticity even after pruning. Evolutionary approaches, such as those proposed by Liquid AI \cite{Hasani2023}, provide automated architecture synthesis, leveraging genetic algorithms to refine network structure dynamically. \\

Reservoir Computing (RC), particularly ESNs, has been a focal point for pruning research. RC is a machine learning approach that leverages the dynamics of a fixed, high-dimensional nonlinear system, known as the reservoir, to process input signals efficiently. For discrete signals, the reservoir state evolution is governed by the map
\begin{equation}
    \mathbf{r}_{t+1} = (1 - \alpha) \mathbf{r}_t + \alpha g(\mathbf{A} \mathbf{r}_t + \mathbf{W}_i \mathbf{u}_t)
    \label{eq:rc}
\end{equation}
where \( \mathbf{r}_t \) represents the reservoir state at time instance \( t \), \( \mathbf{u}_t \) is the input, \( \mathbf{W}_{\mathrm{in}} \) is the input weight matrix, \( g \) is a nonlinear activation function, and \( \alpha \in (0,1] \) is the leakage rate that controls the update speed of the reservoir state. The reservoir, represented by the adjacency matrix  \( \mathbf{A} \), is typically implemented as a randomly connected network. It transforms the input sequence $\mathbf{u}$ into a rich feature representation, while only the optimal linear superposition of those reservoir states $\mathbf{r}$ is trained in the output layer
\begin{equation}
    \mathbf{y}=\mathbf{W}_{\mathrm{out}} \mathbf{r} \quad .
\end{equation}
 
Recent findings in reservoir computing suggest that increasing the size of randomly connected reservoir networks does not always improve performance \cite{EvolveRC_Yadav2025}. Excessive internal interactions can degrade signal propagation, introduce chaotic dynamics, or lead to computational inefficiencies. Studies have shown that large reservoir networks may suffer from signal fading or amplification, reducing their ability to effectively process information \cite{Jaeger2001, Verstraeten2007, Yildiz2012, Gallicchio2017}. These challenges highlight the need for systematic reduction strategies, such as structured pruning, to enhance efficiency and task-specific performance \cite{Lukosevicius2012}. In a larger scope, research on optimal reservoir networks can help to build smaller but higher performing RCs, e.g. for edge computing. \\

Early works \cite{Schrauwen2009} demonstrated that RC models could be pruned effectively without significant loss of computational power. The pruning of ESNs is particularly complex due to the importance of preserving network dynamics, spectral properties, and connectivity structures. Notable research \cite{Lukosevicius2012, Can2024, Schiller2020} has analyzed how different topologies, connectivity distributions, and weight constraints impact pruning outcomes.  \\

Despite advancements, key structural changes during pruning in RC remain poorly understood. Specific open questions include:  
\begin{itemize}  
    \item How do fundamental network properties (e.g., clustering coefficient, spectral radius, average out-degree) evolve with pruning?  
    \item What is the preferred distribution of input-receiving and readout nodes after pruning for optimal task performance?  
    \item How does initial network structure affect long-term stability and generalization in pruned reservoir networks?  
\end{itemize}  

Understanding these structural transformations is essential for refining pruning methods and designing more efficient, task-specific reservoir networks.  \\

By introducing a structured pruning framework, we aim to investigate if the performance of RC can be increased by a performance-informed node removal as well as elucidate changes in the optimal pruned network structure by answering the aforementioned questions. If less nodes allow for more performance, indicators of fundamental principles governing network efficiency are present in the pruned networks and the pruning process. 
%Unlike growth-based strategies, pruning enables the discovery of task-relevant substructures within large networks, offering insights into optimal resource allocation across machine learning, neuroscience, and network science. 
As an alternative to expansion-based optimization, performance-dependent pruning provides a systematic approach to reducing complexity while preserving—or even improving—functional efficiency. We show that for a range of different prediction tasks there \textit{always} exists a reservoir network of same or better performance after pruning certain reservoir nodes.

\section{Methods}
Our analysis is based on the classical reservoir computing paradigm \eqref{eq:rc} using random Erdős–Rényi graphs as reservoir network $\mathbf{A}$ with $N=|\mathbf{A}|$ nodes. Training of the readout matrix is performed via ridge regression with regularization parameter $\lambda$. 
By intention, we set up the RC models such that not all nodes in the reservoir are connected to input and output layers. Specifically, we randomly choose $50\%$ of the reservoir nodes to be input-receiving (at random input weights), and $50\%$ random reservoir nodes to be connected to the trainable read-out layer via a read-out mask. This allows to study how the fraction of input-receiving and output-connected nodes is affected by the pruning, i.e. whether the pruning approach prefers to keep internal or input/output-connected nodes in $\mathbf{A}$. 

The pruning approach in this work is based on iterative and performance-informed node removal from the reservoir network. Starting from the randomly generated ER-graph $\mathbf{A}^{(k=0)}$ of specified size $N_{init}$ and density $\rho_{init}$, a fraction of $f_c=0.25$ reservoir nodes is selected at random to form the set of \textit{candidate} nodes $V_c^{(k)}=\{v_1, \dots, v_c\}$, $c=\lceil p_c \cdot N^{(k)} \rceil $. Independently, $c$ candidate reservoir graphs $\mathbf{A}^{(k)}_{i}$ are generated by removing node $v_i$ from $\mathbf{A}^{(k)}$. Corresponding entries of the read-in weights and the read-out mask are removed accordingly. The resulting RC models are trained, and the mean squared error on the test set is evaluated. The final choice about the node to finally prune from $\mathbf{A}^{(k)}$ in the current iteration $k$ is performance-informed: The node that - when removed - gives the least performance decrease, or the maximal performance increase, will be finally pruned from the reservoir network. The next pruning iteration $k+1$ thus starts from a reservoir network of $|\mathbf{A}^{(k+1)}|=N^{(k)}-1$ nodes, and repeats the performance-informed node removal from newly randomly sampled candidate nodes. \\

Pruning is halted once the performance of the pruned reservoir does not increase (allowing for the patience of five consecutive iterations with performance degradation), or if a minimum number of $N^{(K)}=15$ reservoir nodes is achieved. The model with the lowest test set error from the $K$ pruning iterations is reported as pruned model. The overall approach is greedy, such that the optimal pruning decision is bounded to the current pruning iteration and the set of current candidate nodes. Increasing $f_c$, i.e. the number of candidate nodes per iteration, scales the computational effort, but allows to find better pruning candidates and thus better-pruned models. In the limit of $f_c=1.0$, the truly optimal pruned reservoir network will be obtained in the greedy selection process. Our results display the aggregated results of $50$ pruning trials per experiment, each starting from a randomly generated ER reservoir graph, using RC hyperparameters from a preliminary hyperparameter. The main goal is to identify common properties of pruned models when starting from different initial reservoir graphs with only size, density and spectral radius being initially fixed. Properties on the graph level and on node level are tracked along pruning iterations and across candidate nodes, giving a rich view of what type of high-performing networks arise by pruning what kind of network nodes. All computations were performed using the open-source Python \href{https://github.com/Cyber-Physical-Systems-in-Mech-Eng/pyReCo}{\texttt{pyReCo}} library (developed by the authors) that implements the presented pruning strategy with property logging.    
% \texttt{pyReCo}\footnote{\url{https://github.com/Cyber-Physical-Systems-in-Mech-Eng/pyReCo}}

\section{Results}\label{sec:results}

\begin{figure*}[ht]
\centering
\includegraphics[width=1\textwidth]{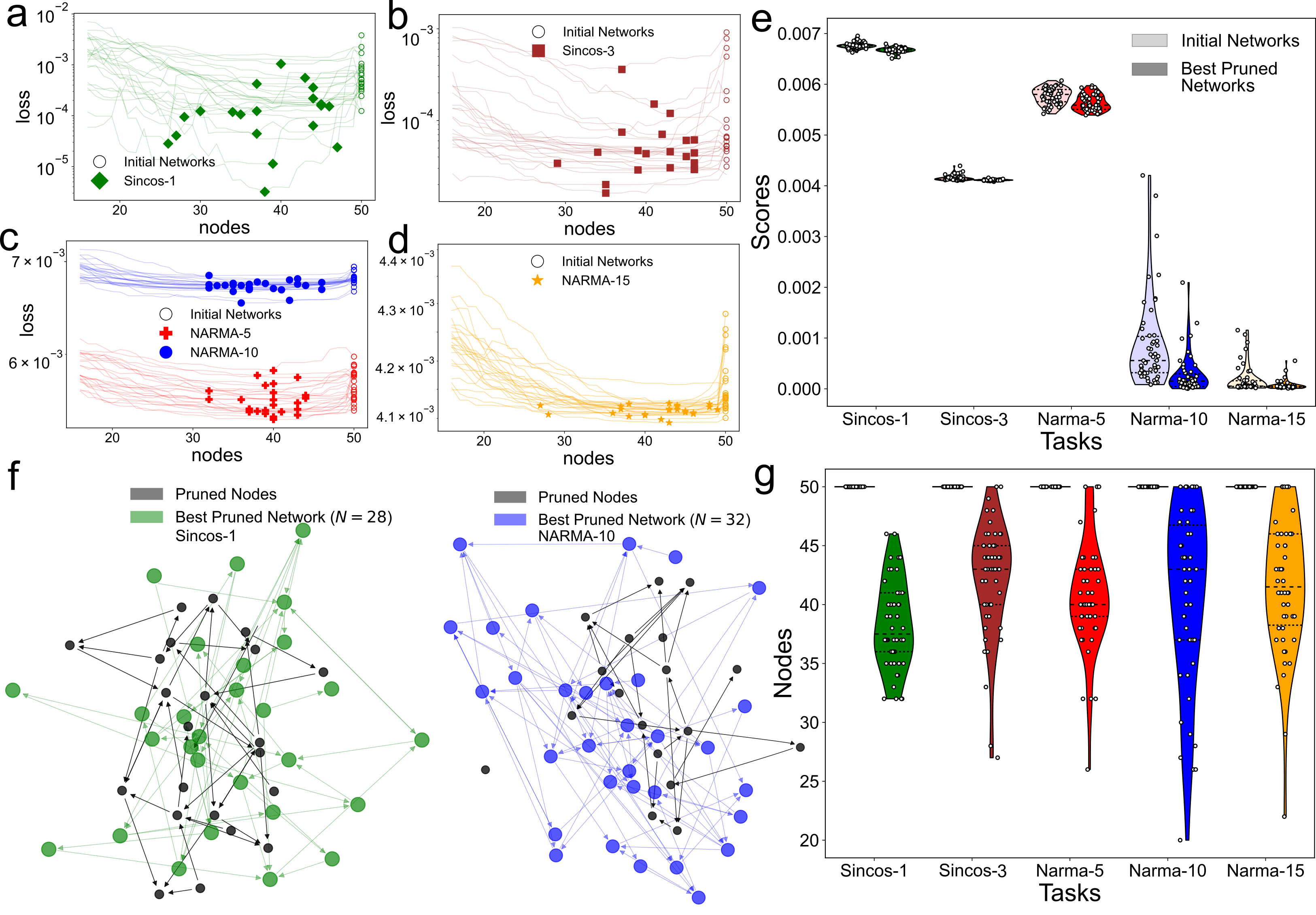}
\caption{\textbf{Node pruning improves efficiency while reducing network size}. The overall loss of the reservoir computers shown with respect to the changing reservoir size as pruning starts from initial Erdős-Rényi random networks (open circles) of size and density, $[N_{init}, \rho{init}] = [50, 0.05]$ for solving (a, b) Sincos-1 and 3, (c)-(d) NARMA-5, 10 and 15 tasks. The solid markers represent the best-performing pruned network. (e) The Mean Square Errors (MSE) of the initial (\textit{left}) and best pruned networks (\textit{right}) are shown for each task. (f) Examplar pruned reservoir networks for Sincos-1 (\textit{left}) and NARMA-10 (\textit{right}) tasks. Pruned nodes are represented with dark grey color. (g) The obtained sizes of best-performing pruned networks (\textit{right}) are compared with the initial networks ($N_{init}=50$) (\textit{left}) for different tasks. 
}\label{fig1}
\end{figure*}

The proposed pruning approach is evaluated on reservoir computers for a set of different sequence-to-sequence learning tasks. First, the memory-free translation of a sinusoidal signal into its corresponding $\pi/4$ phase-shifted copy $\sin\left(t\right)\mapsto \cos\left(t\right)$, or Sinos-1 task. In order to increase complexity, we used a complex version of the Sincos-1 task, where the sine function is mapped to its complex polynomial combination: $\sin\left(t\right) \mapsto \sin\left(t\right)\cos^3\left(t\right)$, namely the Sincos-3 task.\\

Next, we utilized the memory-dependent NARMA$-\tau$ tasks, where $\tau$ indicates the time lag representing the memory of past inputs \cite{Atiya2000}. This is a benchmark task for evaluating a reservoir computer's temporal information processing capability that involves both memory of past inputs and nonlinearity, given by:

\begin{gather}
\begin{aligned}
y(t+1)=\alpha y(t) +\beta y(t)\sum_{i=0}^{M-1}y(t-i) + \\\kappa x(t-M-1)x(t) + \rho
\end{aligned}
\label{eq_narma}
\end{gather}

where $\alpha = 0.3$, $\beta = 0.05$, $\kappa = 1.5$, $\rho = 0.1$ and $M\leq10$. The input $x(t)$ is drawn from a uniform distribution in the
interval $[0,0.5]$. The NARMA function is put inside an additional saturation function $tanh$ for $M>10$, to keep the output bounded between 0 and 1. In this study, we used three NARMA tasks with $\tau=\left\{5, 10, 15\right\}$ cases.
% Lastly, the well known Lorenz (see~\ref{app:lorenz}) and van der Pol (see~\ref{app:vdp}) systems from nonlinear dynamics are studied. 

\subsection{Performance of pruned RCs}

We conducted a systematic investigation of the pruning process across the five aforementioned tasks, recording network performance alongside structural and node-level properties. The initial reservoir networks were constructed as Erdős–Rényi (ER) random graphs with \(N_{\text{init}} = 50\) nodes and an initial density of \(\rho_{\text{init}} = 0.05\). By design, the pruning process led to a sharp decrease in the loss function or a corresponding improvement in network performance. However, the overall performance curve exhibited a distinct trend: as pruning progressed and network size continued to decrease, the loss function initially reached a global minimum before subsequently increasing. This characteristic behavior was consistently observed across all tasks, as illustrated in Fig.~\ref{fig1} (a)-(d). The optimal network structure was identified at this global minimum, corresponding to the highest-performing pruned network. Overall, the best-pruned networks demonstrated superior performance compared to the original ER-random networks, as depicted in the violin plots in Fig.~\ref{fig1} (e). Additionally, Fig.~\ref{fig1} (g) compares the sizes of these optimal pruned networks against their initial counterparts.\\

This initial study confirms that the pruning process effectively enhances the performance of ER-random networks by extracting a more compact yet functionally efficient subnetwork. The pruning mechanism selectively removes nodes that do not contribute to the reservoir’s computational efficacy, resulting in a refined network architecture that optimally processes information. On average, the best-pruned networks comprised \(\lesssim 40\) nodes across all tasks. Notably, the most compact pruned network—obtained for the NARMA-10 task—contained only 20 nodes, representing a \(60\%\) reduction in network size while outperforming its initial ER seed network \(N_{\text{init}} = 50\) size. The pruned nodes primarily included disconnected elements and dead-end nodes, as exemplified in Fig.~\ref{fig1} (f), where removed nodes and their corresponding edges are highlighted in gray. In the subsequent sections, we further analyze the structural and node-level properties of these optimized pruned networks and how they differ from their initial Erdős–Rényi (ER) random seed graphs.

\subsection{Graph-level properties}

\begin{figure*}
\centering
\includegraphics[width=0.95\textwidth]{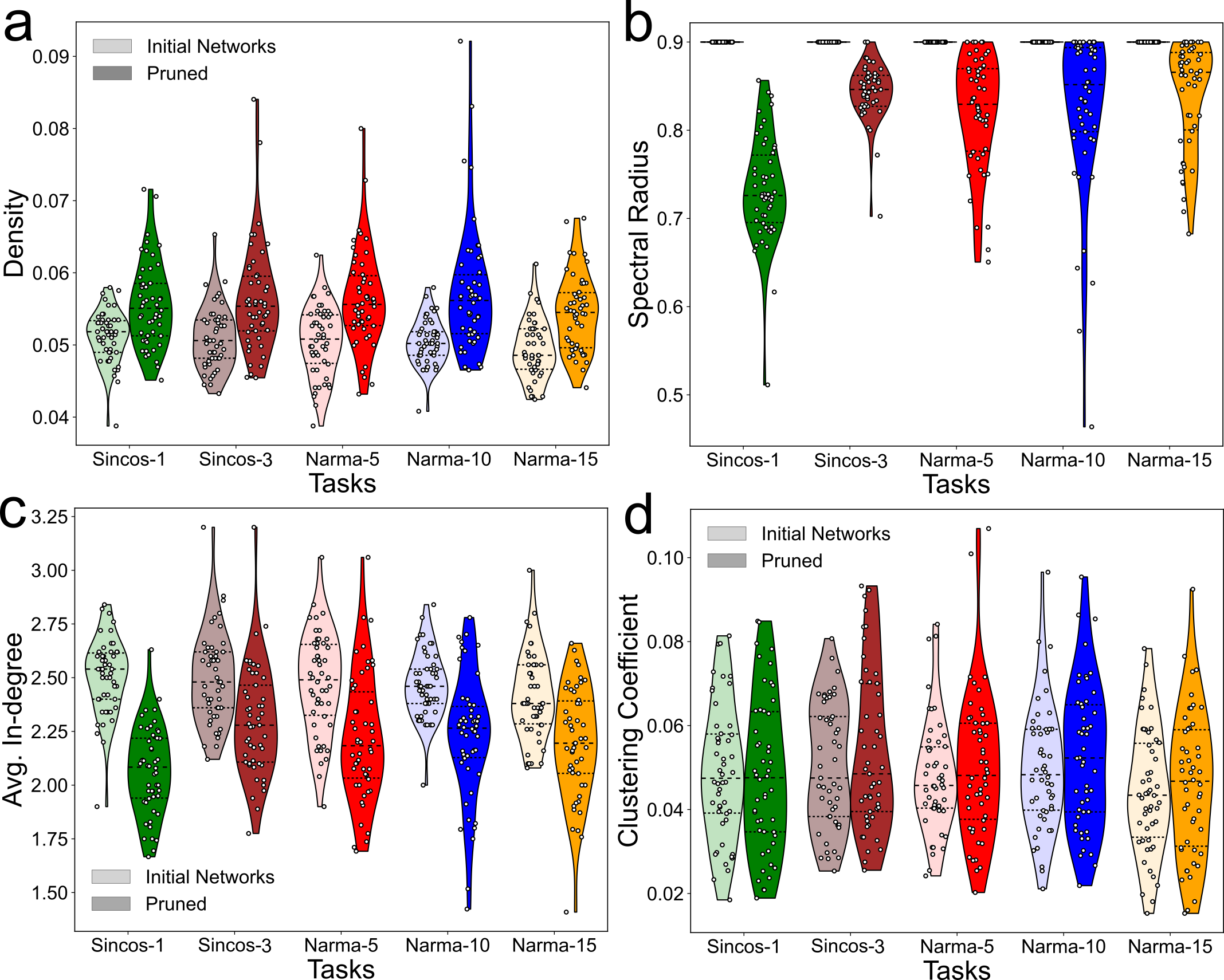}
\caption{\textbf{Change in network structural properties after pruning.} The changes in the reservoir network structure can be easily elucidated using (a) density ($\rho$), (b) spectral radius, (c) average in-degree and (d) clustering coefficient for 5 different tasks. The left and right violins represent the initial and pruned network properties for each task.
\textbf{
}}\label{fig2}
\end{figure*}

% \section{Structural Properties of the Pruned Networks}

To further investigate the structural characteristics of the pruned networks, we analyzed key network properties, namely density, spectral radius, average in-degree, and clustering coefficient. The violin plots in Fig.~\ref{fig2} illustrate the distributions of these properties across the pruning process, capturing their variations between the initial Erdős–Rényi (ER) random networks and the best-pruned networks. These structural measures are crucial for understanding the topological evolution of the reservoir networks as they undergo pruning.

% \subsection{Network Properties and Their Mathematical Definitions}

The \textbf{density} of a directed network is defined as the ratio of existing edges to the maximum possible edges:

\begin{equation}
    \rho = \frac{E}{N(N-1)}
\end{equation}

where \(E\) is the total number of directed edges and \(N\) is the number of nodes in the network. This measure reflects how densely connected the network is. The \textbf{spectral radius} of a network is given by:

\begin{equation}
    \lambda_{\max} = \max |\lambda_i|
\end{equation}

where \(\lambda_i\) are the eigenvalues of the network's adjacency matrix. The spectral radius is crucial in reservoir computing, as it governs the network’s dynamical stability and memory capacity.
Then we calculated the \textbf{average in-degree} of the pruned-networks which is given by:

\begin{equation}
    k_{\text{in,avg}} = \frac{1}{N} \sum_{i=1}^{N} k_{\text{in},i}
\end{equation}

where \(k_{\text{in},i}\) represents the number of incoming connections to node \(i\). This metric provides insight into how information is distributed across the network. And finally, the \textbf{clustering coefficient} that quantifies the tendency of nodes to form local clusters which is defined as:

\begin{equation}
    C = \frac{1}{N} \sum_{i=1}^{N} C_i
\end{equation}

where \(C_i\) is the local clustering coefficient of node \(i\), calculated as:

\begin{equation}
    C_i = \frac{e_i}{k_i(k_i - 1)}
\end{equation}

where \(e_i\) is the number of directed edges between the neighbors of node \(i\), and \(k_i\) is its degree.

% \subsection{Results and Interpretation}

\begin{figure*}
\centering
\includegraphics[width=1\textwidth]{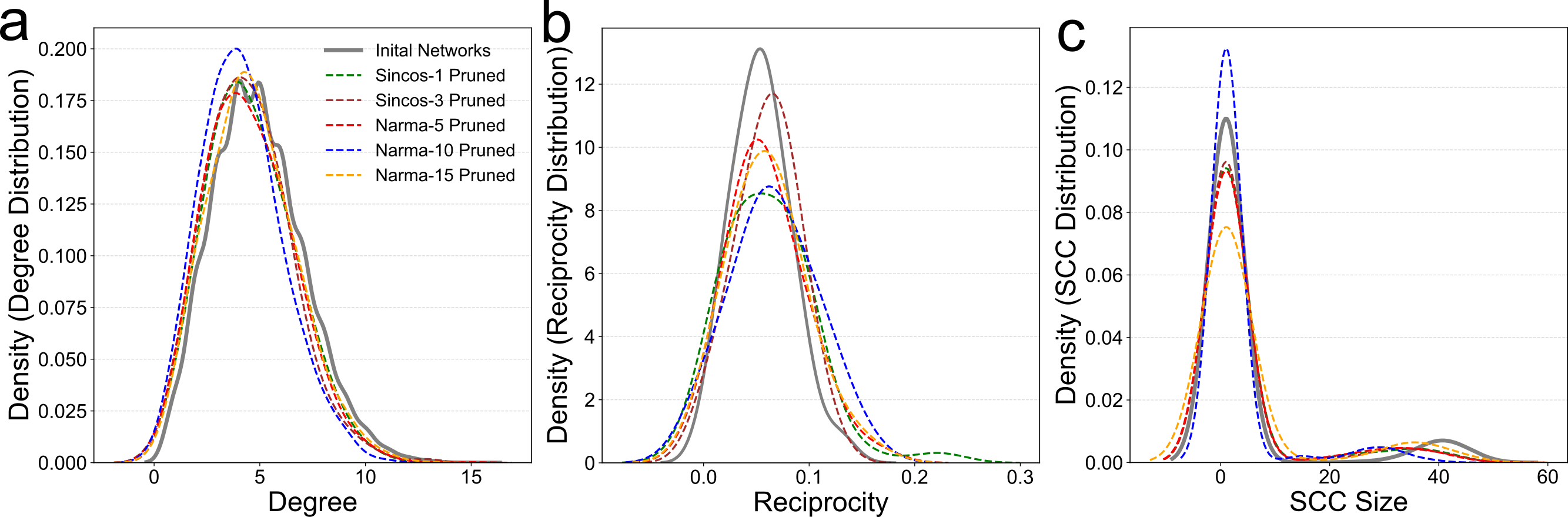}
\caption{\textbf{Degree, Reciprocity, and SCC Distributions:} Kernel Density Estimation (KDE) plots comparing the distributions of (a) degree, (b) reciprocity, and (c) strongly connected components (SCC) for the initial Erdős-Rényi (ER) random networks and the best-performing pruned networks across all cases. The distributions illustrate how pruning alters network connectivity, increasing reciprocity while reducing degree, with varying effects on SCC depending on network complexity.}\label{fig4}
\end{figure*}

The pruning process induced notable structural changes in the networks. As shown in Fig.~\ref{fig2}(a), the density of the best-pruned networks increased across all tasks from the initial mean density of \(\rho_{\text{init}} = 0.05\) to approximately \(0.06\). This increase suggests that the pruning process removes nodes in a way that results in a more compact network, where the remaining nodes retain or even strengthen their connectivity.

A sharp decline in the spectral radius was observed (Fig.~\ref{fig2}(b)). The spectral radius was initially fixed at \(\lambda_{\max} = 0.9\) for all networks to maintain consistency in the initial conditions. However, after pruning, it dropped to values as low as \( < 0.5\). Since the spectral radius governs the stability of dynamical systems, this reduction suggests that pruning systematically eliminates redundant or weakly connected nodes, leading to a network with a more controlled dynamical regime.

The average in-degree of the pruned networks also decreased (Fig.~\ref{fig2}(c)). Since the number of nodes \(N\) decreases during pruning while density \(\rho\) slightly increases, we can express the expected number of incoming connections per node as:

\begin{equation}
    k_{\text{in,avg}} = \rho (N-1).
\end{equation}

Given that \(\rho\) increases modestly while \(N\) drops significantly, the overall trend results in a lower average in-degree. This confirms that pruning removes nodes along with their incoming and outgoing connections, leading to a sparser but more structured network.\\

Interestingly, the clustering coefficient remained largely unchanged throughout pruning (Fig.~\ref{fig2}(d)). This invariance suggests that while individual nodes and edges were removed, the local connectivity patterns—capturing the extent of triangular structures within the network—were preserved. This implies that the pruning mechanism selectively eliminates nodes in a way that does not disrupt local clustering properties, reinforcing the robustness of the network’s mesoscopic organization.\\

\begin{figure*}
\centering
\includegraphics[width=1\textwidth]{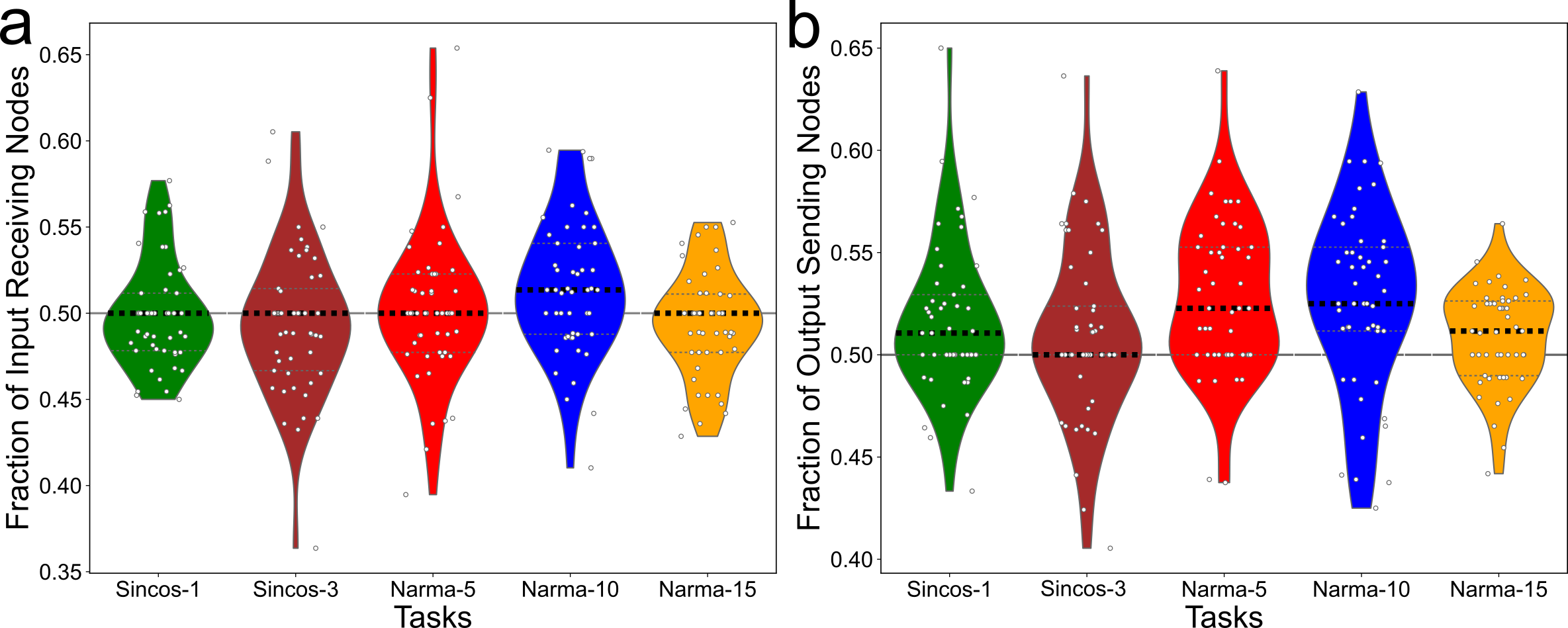}
\caption{
\textbf{Asymmetry between input-receiving and readout nodes.} The fraction of (a) input-receiving and (b) readout nodes, calculated by dividing their number by the total number of pruned network nodes, is shown for various tasks. The grey line in the background indicates their fraction in the initial network, $\mathbf{A}^{(k=0)}$.
}\label{fig3}
\end{figure*}

Overall, these structural changes highlight the transformation of the Erdős–Rényi (ER) random networks' macroscopic properties into a more structured, task-optimized network through pruning. In its initial state, the ER network exhibits a homogeneous connectivity distribution with a random arrangement of nodes and edges. However, the pruning process selectively eliminates redundant and weakly connected nodes, leading to a refined network with increased density, reduced spectral radius, and lower average in-degree, all while preserving local clustering properties. This transition suggests that pruning extracts a more functionally efficient core network while discarding structurally insignificant components.  

These macroscopic structural changes provide insights into the global evolution of the pruned networks, they explain the underlying mechanisms driving this transformation to a greater extent. Furthermore, to gain a deeper understanding, it is essential to analyze node-level properties, which reveal how individual nodes contribute to the emergent computational efficiency of the pruned network. In the following section, we examine these node-level characteristics to uncover the finer details of the network’s reorganization and the roles played by different types of nodes in optimizing reservoir computing performance.

\subsection{Beneficial node-level properties}

We calculated node-level properties of the initial and finally obtained best-pruned networks in order to understand the substructural changes in the network when the nodes are selectively removed while improving the performance. The key properties analyzed include degree, reciprocity, and strongly connected components (SCC).

Reciprocity measures the symmetry of connections in a network and is defined as:

\[
R = \frac{\sum_{i,j} A_{ij} A_{ji}}{\sum_{i,j} A_{ij}}.
\]

As shown in Table \ref{tab:network_properties}, pruning generally improves reciprocity across the cases, with increases of \textbf{11.51\%} in Sincos-1, \textbf{11.68\%} in Sincos-3, and \textbf{6.49\%} in Narma-5. This suggests more balanced node interactions post-pruning. In Narma-10 and Narma-15, the increase is more significant, with improvements of \textbf{14.00\%} and \textbf{17.89\%}, respectively, indicating that pruning leads to a more efficient and balanced structure, especially in more complex networks.\\

Degree, representing the number of connections each node has, generally decreases after pruning, as seen in Table \ref{tab:network_properties}. The largest reduction occurs in Narma-10 with a \textbf{17.27\%} decrease, reflecting a more significant pruning effect on connectivity in larger networks. Other cases, such as Sincos-1 (\textbf{-7.82\%}) and Narma-5 (\textbf{-9.30\%}), show a decrease in degree, suggesting that pruning simplifies the network structure by reducing the number of connections. This may contribute to improved network performance by focusing on the most essential connections. This trend is further supported by the KDE (Kernel Density Estimation) plots shown in Fig.\ref{fig4}, which illustrate how the degree distribution shifts between the initial and pruned networks. The KDE curves reveal a noticeable narrowing of the degree distribution after pruning, indicating that pruning focuses on a more limited set of connections, which aligns with the degree reduction observed.\\

Strongly Connected Components (SCC) measure the network's cohesiveness, with higher values indicating better robustness. SCC is defined as:

\[
\text{SCC} = \sum_{i} \sum_{j} \delta(A_{ij}, A_{ji}),
\]

where \( \delta(x, y) \) measures directed cycles. As shown in Table \ref{tab:network_properties}, pruning tends to reduce SCC in smaller networks, such as Sincos-1 (\textbf{-12.48\%}) and Narma-5 (\textbf{-4.25\%}). However, in more complex cases like Narma-10, SCC increases by \textbf{20.85\%}, suggesting that pruning can strengthen the cohesiveness of larger networks by eliminating weaker connections and reinforcing stronger ones. The KDE plot in Fig.\ref{fig4} also reveals the shift in SCC distribution post-pruning. It shows that the distribution for the pruned networks tends to be more concentrated around higher values in some cases, especially in larger networks, corresponding to the observed increase in SCC for Narma-10.\\

\begin{table*}[ht]
    \centering
    \begin{tabular}{|l|l|l|l|l|l|l|}
    \hline
    \textbf{Case} & \textbf{Degree (Init/Final)} & \textbf{\% Change} & \textbf{Reciprocity} & \textbf{\% Change} & \textbf{SCC} & \textbf{\% Change} \\
    \hline
    Sincos-1 & 4.932 / 4.546 & -7.82\% & 0.0548 / 0.0611 & 11.51\% & 11.54 / 10.10 & -12.48\% \\
    \hline
    Sincos-3 & 4.827 / 4.446 & -7.90\% & 0.0537 / 0.0599 & 11.68\% & 11.66 / 10.58 & -9.26\% \\
    \hline
    Narma-5  & 4.944 / 4.484 & -9.30\% & 0.0560 / 0.0597 & 6.49\%  & 10.36 / 9.92  & -4.25\% \\
    \hline
    Narma-10 & 5.015 / 4.149 & -17.27\% & 0.0582 / 0.0663 & 14.00\% & 10.36 / 12.52 & 20.85\% \\
    \hline
    Narma-15 & 4.994 / 4.624 & -7.40\% & 0.0514 / 0.0606 & 17.89\% & 10.08 / 8.41  & -16.60\% \\
    \hline
    \end{tabular}
    \caption{The percent change between the initial and pruned network properties.}
    \label{tab:network_properties}
\end{table*}

Upon further analysis, we identified an emergent node-specific characteristic in the pruned networks: a self-organized asymmetric distribution of input-receiving and readout nodes. Initially, in all our simulations, nodes in the Erdős–Rényi (ER) random networks were randomly designated as input-receiving and readout nodes with an independent probability of 50\%. As a result, the initial reservoir networks, on average, had half of their nodes assigned as input-receiving and half as readout nodes, with approximately 25\% of the total nodes serving both roles.  

However, in the final pruned networks, this initially symmetric distribution became asymmetric. Across all tasks, the fraction of readout nodes consistently exceeded 0.5 in the pruned networks, as illustrated in Fig.~\ref{fig3}. Surprisingly, the proportion of input-receiving nodes remained at 0.5 even after pruning. This indicates that nodes not serving as readout nodes were preferentially removed, suggesting that the pruned networks enhance efficiency by selectively retaining more readout nodes than input nodes while reducing overall network size.  
\\

In summary, as depicted in Table \ref{tab:network_properties} and visualized in the KDE plots (Fig.\ref{fig4}), pruning tends to simplify networks by reducing degrees and increasing reciprocity, with mixed effects on SCC depending on network complexity. Smaller networks generally show reduced SCC after pruning, while larger networks may benefit from stronger connectivity and improved cohesion. This indicates that pruning not only reduces the size of the network but also enhances its structural efficiency. Additionally, pruning leads to a self-organized asymmetric distribution of input-receiving and readout nodes. While the initial networks have a symmetric assignment, with approximately 50\% of nodes designated as input-receiving and 50\% as readout nodes, the final pruned networks consistently retain a higher fraction of readout nodes while maintaining the input-receiving fraction at 0.5. This suggests that non-readout nodes are selectively removed, allowing the pruned networks to maintain efficiency with a reduced structure while prioritizing readout functionality.  

\begin{figure*}
\centering
\includegraphics[width=1\textwidth]{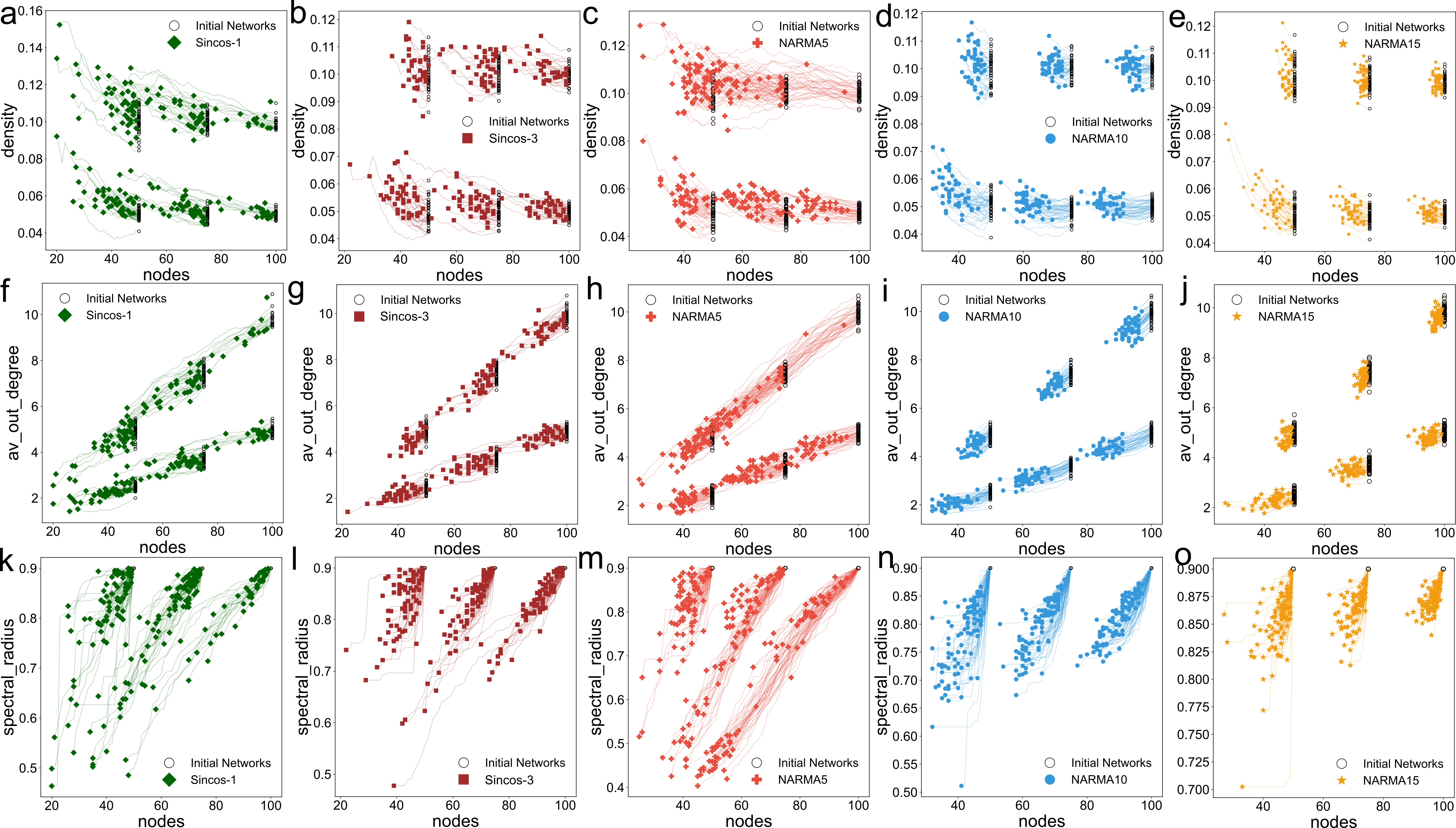}
\caption{
\textbf{Effect of initial network organization on pruned substructure.} Inital Erdős-Rényi random reservoir networks $\mathbf{A}^{(k=0)}$ are generated to have different sizes $N_{init}=[50, 75, 100]$ and densities $\rho_{init}=[0.05, 0.1]$. These are pruned for 5 different Sincos and NARMA tasks. The changes in their densities (\textit{first row}), average out-degree (\textit{middle row}) and spectral radius (\textit{last row}) are shown in different panels. The initial networks are shown with open black circles and the best pruned networks with solid symbols. The pruning trajectories (faded lines) are shown here until the best pruned networks.}\label{fig5}
\end{figure*}

\subsection{Effect of Initial Network Conditions on Pruned Substructures}

In this section, we explore the effect of initial conditions on pruned substructures by testing different Erdős-Rényi random networks for sizes [50, 75, 100] and densities [0.05, 0.1]. The results, summarized in Table \ref{table:reservoir_network_init} and illustrated in Figure \ref{fig5}, highlight how these initial configurations influence reservoir properties such as spectral radius, average in-degree, average out-degree, and clustering coefficient (CC) after pruning.

Smaller, sparser networks (50 nodes, density 0.05) experience substantial reductions in network properties, particularly spectral radius and degree metrics, compared to larger, denser networks (75 and 100 nodes, density 0.1). The spectral radius, average in-degree, out-degree, and CC all decrease significantly in these smaller networks, suggesting that they are more sensitive to pruning. In contrast, larger and denser networks are more resilient, showing smaller changes in their structural properties after pruning, indicating that they are better able to maintain their core structure.

For task-specific results, the SinCos-1 task with 75 nodes and a density of 0.1 shows a 10.67\% decrease in CC after pruning. In contrast, Narma-5 with 100 nodes and a density of 0.1 experiences minimal changes in CC, with a slight increase observed post-pruning. This reflects the greater stability of larger networks. For more complex tasks like Sincos-3, pruning leads to a notable increase in CC when pruning networks with 50 nodes and a density of 0.05. Conversely, Narma-10 with 50 nodes and a density of 0.05 experiences a 23.22\% decrease in CC, illustrating the typical reduction in connectivity for smaller networks. Narma-15 with 75 nodes and a density of 0.1 shows a slight increase in CC (0.16\%) post-pruning, indicating some variation in pruning effects even within larger networks. These task-specific results emphasize how network size and task complexity interact to influence the impact of pruning.\\

The \textbf{spectral radius}, initially fixed at 0.9, is a key indicator of network stability and dynamics. Across tasks, pruning leads to a decrease in spectral radius, with smaller networks experiencing more significant reductions. For SinCos-1, pruning reduces the spectral radius by 10.36\% for 50 nodes (density 0.05) and by 5.71\% for 100 nodes (density 0.1). This suggests that pruning reduces the network's ability to maintain stable dynamics, especially in smaller networks. In Sincos-3, pruning causes a smaller decrease in spectral radius compared to SinCos-1, with a 4.22\% decrease for 50 nodes (density 0.05) and a 4.01\% decrease for 100 nodes (density 0.1), indicating that larger networks are more stable post-pruning. For Narma-5, the spectral radius decreases moderately, while in more complex tasks like Narma-10, pruning results in a significant decrease in spectral radius, particularly for smaller networks (17.07\% for 50 nodes and density 0.05). This highlights the greater impact of pruning on smaller networks and tasks requiring higher dynamical capacity. Larger networks, such as Narma-15, show more stability post-pruning, with a smaller decrease (5.26\%) observed for 100 nodes (density 0.1).

As seen in Fig.\ref{fig5} (last row), the spectral radius always drastically decreases while pruning the network, especially for smaller networks. This suggests that smaller networks, particularly those with lower density, are more likely to lose their ability to maintain stable dynamical behavior. The initial spectral radius of 0.9 ensures that all networks start with a stable dynamical range, and the observed reductions in spectral radius, particularly for tasks like Narma-10, highlight the challenges of preserving dynamical properties after pruning. However, larger networks demonstrate greater stability, indicating that pruning has a lesser impact on their spectral radius and overall stability.

In terms of the \textbf{average in-degree}, pruning generally reduces the connectivity of the network. For SinCos-1 with 50 nodes and density 0.05, the average in-degree decreases by 8.54\%, while for Narma-5 with 100 nodes and density 0.1, the reduction is 6.77\%. The reduction in connectivity is more pronounced in smaller networks, indicating that pruning tends to remove less connected nodes, which affects the network’s capacity to propagate information. In more complex tasks like Narma-10, pruning leads to an 18.15\% decrease in average in-degree for 50 nodes (density 0.05), demonstrating the larger impact pruning has on smaller networks.

Finally, the \textbf{clustering coefficient} (CC) shows varied changes across tasks and network sizes. Generally, pruning leads to a decrease in CC, particularly in smaller networks. For instance, in SinCos-1 with 75 nodes and a density of 0.1, the CC decreases by 10.67\%, while for Narma-5 with 100 nodes and a density of 0.1, a slight increase of 0.16\% is observed. In Sincos-3, pruning causes a significant increase in CC for 50 nodes and a density of 0.05 (50.38\%), suggesting that for some tasks and configurations, pruning can enhance local clustering. In contrast, Narma-10 shows a typical decrease in CC post-pruning, especially in smaller networks, highlighting the varying effects pruning has depending on the task complexity and network size.\\

In conclusion, pruning tends to reduce network stability, as evidenced by decreases in spectral radius, average in- and out-degree, and clustering coefficient, particularly in smaller networks. Larger networks, however, show greater resilience, with smaller decreases in spectral radius and clustering coefficient. Tasks like Narma-5 and Sincos-3 show a more stable network structure post-pruning, while tasks like Narma-10 experience greater reductions in connectivity and dynamical capacity. These findings emphasize the need to carefully consider initial network conditions and task complexity when designing efficient reservoir networks.

\section{Discussion and Outlook}\label{sec:discussion}

This study systematically investigates the structural transformation of Erd\H{o}s--R\'enyi (ER) random networks through a performance-dependent pruning process in the context of reservoir computing (RC). Our results demonstrate that pruning selectively eliminates redundant nodes while preserving key structural properties essential for effective information processing. This process results in a refined network with increased density, reduced spectral radius, and a lower average in-degree while maintaining the clustering coefficient. The observed improvements in network performance highlight the emergence of a compact, efficient computational structure from an initially random topology.

The findings have significant implications for Reservoir Computing (RC), Network Science, and Nonlinear Dynamics communiy where understanding the evolution of complex networks is crucial. In RC, an optimized network structure enhances memory capacity and prediction accuracy, making it beneficial for time-series forecasting, signal processing, and dynamical system modeling~\cite{Jaeger2001, Maass2002}. In \textbf{network science}, this work sheds light on how connectivity patterns evolve through optimization mechanisms, which has implications for communication networks, social networks, and artificial intelligence architectures. In \textbf{nonlinear dynamics}, where RC has been widely used to model chaotic systems, this study contributes a new theoretical perspective on how pruning affects dynamical properties~\cite{Pathak2018, Gauthier2021}.

Beyond artificial networks, \textbf{biological systems} exhibit similar pruning phenomena, reinforcing the broader applicability of the framework introduced in this study. In 	\textbf{neuronal networks}, synaptic pruning plays a fundamental role in cognitive development, where unnecessary synaptic connections are eliminated to optimize brain function~\cite{Huttenlocher1979, Chechik1998}. This process mirrors our observation that pruning reduces network size while improving performance, as seen in the decline and subsequent optimization of the loss function. The removal of redundant nodes in our networks parallels the way the brain refines its connectivity to enhance computational efficiency.

Similarly, in \textbf{gene regulatory networks}, non-essential genes or regulatory interactions are selectively silenced over evolutionary time to improve the efficiency of cellular responses~\cite{Alon2006}. Our results show a drop in spectral radius and a decline in average in-degree, suggesting that removing weakly connected nodes improves the network's functional efficiency—just as biological systems streamline regulatory interactions to optimize control over gene expression. In \textbf{protein interaction networks}, certain interactions are lost as a system adapts to environmental changes, leading to optimized metabolic and signaling pathways~\cite{Barabasi2004}. The increase in density observed in our pruned networks suggests that, despite a reduction in size, the retained nodes maintain or strengthen their functional interconnectivity, similar to how essential protein interactions persist even as redundant ones disappear.

From a real-world perspective, pruning-like mechanisms are observed in various infrastructural and technological networks. In \textbf{communication networks}, inefficient nodes or links are decommissioned to improve efficiency, much like our study's pruning process removes underperforming nodes while enhancing the reservoir’s predictive capabilities. In \textbf{power grids}, redundant connections are removed to reduce energy losses while maintaining resilience, similar to how pruning preserves network connectivity despite node removal. Likewise, in \textbf{transportation systems}, route optimizations eliminate underused pathways while maintaining network connectivity—aligning with our observation that clustering coefficients remain largely unchanged, ensuring the pruned network retains its essential structural motifs.

These parallels suggest that the structural changes induced by pruning in our study align with fundamental optimization strategies observed across biological and engineered systems. By systematically quantifying how pruning refines network efficiency, our work provides a theoretical foundation to understand how complex systems evolve toward more efficient forms.

This work establishes a theoretical foundation for performance-dependent pruning in reservoir computing and complex networks, opening several avenues for future research. Given the parallels with neuronal pruning, this approach could be tested on biologically inspired spiking neural networks to explore how optimized reservoir structures emerge naturally in the brain. Furthermore, the insights from this study can be extended to fields such as \textbf{epidemiology} (modeling how redundant connections in disease-spread networks affect transmission dynamics), \textbf{financial networks} (identifying critical nodes in economic systems), and \textbf{AI explainability} (understanding how deep networks can be pruned without loss of function).\\

Reservoir computing (RC) has gained significant attention for its ability to efficiently process temporal data with minimal training overhead~\cite{Jaeger2001, Maass2002}. While much research in the RC community has focused on optimizing hyperparameters~\cite{Verstraeten2007}, spectral properties~\cite{Rodan2011}, and learning rules~\cite{Schrauwen2008}, little attention has been paid to the structural efficiency of reservoir networks. Most studies employ standard random networks (Erd\H{o}s--R\'enyi, scale-free, or small-world) without systematically analyzing whether these topologies are the most efficient for a given task. Our study addresses this gap by demonstrating that an initially unoptimized random network can be pruned into a more efficient structure without loss of performance, and often with performance enhancement.

Another largely unexplored direction in RC is using it as a framework to study network evolution and pruning mechanisms. While network pruning is well known in deep learning~\cite{LeCun1990, Han2015}, its role in recurrent architectures like RC remains underexplored. By leveraging pruning, we show that redundant structures in reservoir networks can be systematically removed to obtain a more efficient computational core, making RC not only a powerful tool for learning but also a framework for studying the emergent properties of complex systems.

This study also complements the evolving network approach in reservoir computing proposed by Yadav et al.\cite{EvolveRC_Yadav2025}, which takes a \textit{bottom-up} perspective by building performance-dependent networks from scratch through an evolutionary process. Their work focuses on how networks can be grown adaptively to optimize performance, whereas this study follows a \textit{top-down} approach, starting from an initially large network and systematically reducing it to its most efficient form. Together, these two approaches provide a holistic view of how networks can be both grown and pruned to achieve optimal computational structures.

A key insight from this work is that network pruning can reveal the essential structural motifs required for optimal RC performance, while Yadav et al.’s approach demonstrates how these structures can emerge through adaptive growth. Future research could integrate these two methodologies—combining evolutionary network growth with performance-dependent pruning—to refine reservoir topologies dynamically. Such an approach could be particularly valuable in adaptive learning systems, where networks continuously evolve in response to changing tasks and environments.

By bridging insights from network science, machine learning, and neuroscience, this work opens new avenues for understanding the interplay between structure and function in complex adaptive systems.

\begin{acknowledgments}
This work was supported by the Deutsche Forschungsgemeinschaft (DFG, German Research Foundation) under the Special Priority Program (SPP 2353: Daring More Intelligence – Design Assistants in Mechanics and Dynamics, project number 501847579). 
\end{acknowledgments}

\section*{Data Availability Statement}

All the Python codes for the numerical simulations and the data generated in this study are publicly available at the following repository: \href{https://github.com/Cyber-Physical-Systems-in-Mech-Eng/PrunedRC}{https://github.com/Cyber-Physical-Systems-in-Mech-Eng/PrunedRC}.

\appendix

\section{Quantification of the effect of different initial conditions}

The effect of different initial Erods-Renyi random networks on the emergent best-pruuned subnetwork properties are provided in the following Table \ref{table:reservoir_network_init}. A combination of different nodes $N_{init}=[50, 75, 100]$ and densities $\rho_{init}=[0.05, 0.1]$ are used used to generate a total of six different kinds of initial ER-random networks $\mathbf{A}^{(k=0)}$ for pruning experiment for each of the 5 tasks. The table shows the mean percent change of density, spectral radius, average in-degree, average out-degree and clustering coefficient (CC) of the final pruned networks from that of the initial networks.

\begin{table*}[ht]
    \centering
    \begin{tabular}{|c|c|c|c|c|c|c|c|}
    \hline
    \textbf{Tasks} & $N_{init}$ & $\rho_{init}$ & \textbf{$\Delta \rho$ (\%)} & \textbf{$\Delta$ Spectral radius (\%)} & \textbf{$\Delta$ Avg. in-degree (\%)} & \textbf{$\Delta$ Avg. out-degree (\%)} & \textbf{$\Delta$ CC (\%)} \\
    \hline
    SinCos-1 & 50 & 0.05 & 4.23\% & -10.36\% & -8.54\% & -8.54\% & -2.41\% \\
    SinCos-1 & 50 & 0.1  & 0.73\% & -4.18\%  & -5.44\% & -5.44\% & 3.03\% \\
    SinCos-1 & 75 & 0.05 & 0.60\% & -1.80\%  & -2.11\% & -2.11\% & -10.67\% \\
    SinCos-1 & 75 & 0.1  & 0.11\% & -7.84\%  & -6.65\% & -6.65\% & 0.62\% \\
    SinCos-1 & 100 & 0.05 & 2.77\% & -0.66\%  & -0.34\% & -0.34\% & 0.84\% \\
    SinCos-1 & 100 & 0.1  & 1.20\% & -5.71\%  & -5.95\% & -5.95\% & 3.90\% \\
    \hline
    SinCos-3 & 50 & 0.05 & 9.53\% & -4.22\%  & -8.35\% & -8.35\% & 50.38\% \\
    SinCos-3 & 50 & 0.1  & -0.23\% & -11.98\% & -10.41\% & -10.41\% & -4.90\% \\
    SinCos-3 & 75 & 0.05 & 3.67\% & -11.28\% & -10.34\% & -10.34\% & -5.58\% \\
    SinCos-3 & 75 & 0.1  & 2.18\% & -2.17\%  & -1.96\% & -1.96\% & -1.00\% \\
    SinCos-3 & 100 & 0.05 & 2.23\% & -0.67\%  & -0.86\% & -0.86\% & -2.77\% \\
    SinCos-3 & 100 & 0.1  & 1.12\% & -4.01\%  & -3.99\% & -3.99\% & 1.70\% \\
    \hline
    Narma-5  & 50 & 0.05 & 0.65\% & -16.12\% & -13.73\% & -13.73\% & -4.01\% \\
    Narma-5  & 50 & 0.1  & 1.73\% & 0.00\%   & -0.34\% & -0.34\% & 2.30\% \\
    Narma-5  & 75 & 0.05 & 3.79\% & -11.25\% & -11.64\% & -11.64\% & -3.20\% \\
    Narma-5  & 75 & 0.1  & 0.90\% & -2.41\%  & -1.83\% & -1.83\% & -1.89\% \\
    Narma-5  & 100 & 0.05 & 6.09\% & -7.41\%  & -6.77\% & -6.77\% & 0.16\% \\
    Narma-5  & 100 & 0.1  & 8.44\% & -11.53\% & -11.28\% & -11.28\% & 3.87\% \\
    \hline
    Narma-10 & 50 & 0.05 & 8.39\% & -17.07\% & -18.15\% & -18.15\% & -23.22\% \\
    Narma-10 & 50 & 0.1  & 2.35\% & -13.02\% & -10.18\% & -10.18\% & 1.67\% \\
    Narma-10 & 75 & 0.05 & 4.14\% & -19.39\% & -16.97\% & -16.97\% & 22.05\% \\
    Narma-10 & 75 & 0.1  & 0.29\% & -3.72\%  & -3.77\% & -3.77\% & -1.47\% \\
    Narma-10 & 100 & 0.05 & -0.18\% & -12.54\% & -11.27\% & -11.27\% & -7.67\% \\
    Narma-10 & 100 & 0.1  & -0.91\% & -10.00\% & -9.92\% & -9.92\% & -0.96\% \\
    \hline
    Narma-15 & 50 & 0.05 & 4.38\% & -3.93\%  & -4.14\% & -4.14\% & -10.15\% \\
    Narma-15 & 50 & 0.1  & 2.29\% & -1.72\%  & -1.89\% & -1.89\% & 1.05\% \\
    Narma-15 & 75 & 0.05 & 0.39\% & -3.67\%  & -5.04\% & -5.04\% & -6.38\% \\
    Narma-15 & 75 & 0.1  & -0.36\% & -3.32\%  & -3.06\% & -3.06\% & 0.16\% \\
    Narma-15 & 100 & 0.05 & 2.73\% & -5.26\%  & -6.60\% & -6.60\% & -0.03\% \\
    Narma-15 & 100 & 0.1  & 0.23\% & -2.89\%  & -3.82\% & -3.82\% & 1.60\% \\
    \hline
    \end{tabular}
    \caption{The changes in reservoir network properties for different initial Erdős-Rényi random networks.}
    \label{table:reservoir_network_init}
\end{table*}

\section{Pruning model parameters}

The following parameters are fixed and used for all the experiments in this study in order to maintain consistency: Minimun nodes till the pruning continues, $N^{K}=15$, patience (=5) defines the continuation of pruning process even after reaching a (local) minimum of the test set score, fraction of input-receiving and readout nodes is 0.25, set of candidate nodes used to find a node to be ruined, $f_{c}$=0.25, spectral radius = 0.9, mean square error (MSE) is used for obtaining the performance.

\nocite{*}
% \bibliography{aipsamp}% Produces the bibliography via BibTeX.

\begin{thebibliography}{38}%
\makeatletter
\providecommand \@ifxundefined [1]{%
 \@ifx{#1\undefined}
}%
\providecommand \@ifnum [1]{%
 \ifnum #1\expandafter \@firstoftwo
 \else \expandafter \@secondoftwo
 \fi
}%
\providecommand \@ifx [1]{%
 \ifx #1\expandafter \@firstoftwo
 \else \expandafter \@secondoftwo
 \fi
}%
\providecommand \natexlab [1]{#1}%
\providecommand \enquote  [1]{``#1''}%
\providecommand \bibnamefont  [1]{#1}%
\providecommand \bibfnamefont [1]{#1}%
\providecommand \citenamefont [1]{#1}%
\providecommand \href@noop [0]{\@secondoftwo}%
\providecommand \href [0]{\begingroup \@sanitize@url \@href}%
\providecommand \@href[1]{\@@startlink{#1}\@@href}%
\providecommand \@@href[1]{\endgroup#1\@@endlink}%
\providecommand \@sanitize@url [0]{\catcode `\\12\catcode `\$12\catcode `\&12\catcode `\#12\catcode `\^12\catcode `\_12\catcode `\%12\relax}%
\providecommand \@@startlink[1]{}%
\providecommand \@@endlink[0]{}%
\providecommand \url  [0]{\begingroup\@sanitize@url \@url }%
\providecommand \@url [1]{\endgroup\@href {#1}{\urlprefix }}%
\providecommand \urlprefix  [0]{URL }%
\providecommand \Eprint [0]{\href }%
\providecommand \doibase [0]{http://dx.doi.org/}%
\providecommand \selectlanguage [0]{\@gobble}%
\providecommand \bibinfo  [0]{\@secondoftwo}%
\providecommand \bibfield  [0]{\@secondoftwo}%
\providecommand \translation [1]{[#1]}%
\providecommand \BibitemOpen [0]{}%
\providecommand \bibitemStop [0]{}%
\providecommand \bibitemNoStop [0]{.\EOS\space}%
\providecommand \EOS [0]{\spacefactor3000\relax}%
\providecommand \BibitemShut  [1]{\csname bibitem#1\endcsname}%
\let\auto@bib@innerbib\@empty
%</preamble>
\bibitem [{\citenamefont {Selemon}(2013)}]{Selemon2013}%
  \BibitemOpen
  \bibfield  {author} {\bibinfo {author} {\bibfnamefont {L.~D.}\ \bibnamefont {Selemon}},\ }\bibfield  {title} {\enquote {\bibinfo {title} {A role for synaptic plasticity in the adolescent development of executive function},}\ }\href {\doibase https://www.nature.com/articles/tp2013184} {\bibfield  {journal} {\bibinfo  {journal} {Translational Psychiatry}\ }\textbf {\bibinfo {volume} {3}},\ \bibinfo {pages} {e238} (\bibinfo {year} {2013})}\BibitemShut {NoStop}%
\bibitem [{\citenamefont {Barabási}\ and\ \citenamefont {Oltvai}(2004)}]{Barabasi2004}%
  \BibitemOpen
  \bibfield  {author} {\bibinfo {author} {\bibfnamefont {A.-L.}\ \bibnamefont {Barabási}}\ and\ \bibinfo {author} {\bibfnamefont {Z.~N.}\ \bibnamefont {Oltvai}},\ }\bibfield  {title} {\enquote {\bibinfo {title} {Network biology: Understanding the cell’s functional organization},}\ }\href {\doibase 10.1038/nrg1272} {\bibfield  {journal} {\bibinfo  {journal} {Nature Reviews Genetics}\ }\textbf {\bibinfo {volume} {5}},\ \bibinfo {pages} {101--113} (\bibinfo {year} {2004})}\BibitemShut {NoStop}%
\bibitem [{\citenamefont {Alon}(2007)}]{Alon2007}%
  \BibitemOpen
  \bibfield  {author} {\bibinfo {author} {\bibfnamefont {U.}~\bibnamefont {Alon}},\ }\href {\doibase https://www.taylorfrancis.com/books/mono/10.1201/9781420011432/introduction-systems-biology-uri-alon} {\emph {\bibinfo {title} {An Introduction to Systems Biology: Design Principles of Biological Circuits}}}\ (\bibinfo  {publisher} {Chapman \& Hall/CRC},\ \bibinfo {year} {2007})\BibitemShut {NoStop}%
\bibitem [{\citenamefont {Sporns}\ and\ \citenamefont {Betzel}(2016)}]{Sporns2016}%
  \BibitemOpen
  \bibfield  {author} {\bibinfo {author} {\bibfnamefont {O.}~\bibnamefont {Sporns}}\ and\ \bibinfo {author} {\bibfnamefont {R.~F.}\ \bibnamefont {Betzel}},\ }\bibfield  {title} {\enquote {\bibinfo {title} {Modular brain networks},}\ }\href {\doibase 10.1146/annurev-psych-122414-033634} {\bibfield  {journal} {\bibinfo  {journal} {Annual Review of Psychology}\ }\textbf {\bibinfo {volume} {67}},\ \bibinfo {pages} {613--640} (\bibinfo {year} {2016})}\BibitemShut {NoStop}%
\bibitem [{\citenamefont {Bellec}\ \emph {et~al.}(2018)\citenamefont {Bellec}, \citenamefont {Scherr}, \citenamefont {Subramoney}, \citenamefont {Hajek}, \citenamefont {Salaj}, \citenamefont {Legenstein},\ and\ \citenamefont {Maass}}]{Bellec2018}%
  \BibitemOpen
  \bibfield  {author} {\bibinfo {author} {\bibfnamefont {G.}~\bibnamefont {Bellec}}, \bibinfo {author} {\bibfnamefont {F.}~\bibnamefont {Scherr}}, \bibinfo {author} {\bibfnamefont {A.}~\bibnamefont {Subramoney}}, \bibinfo {author} {\bibfnamefont {E.}~\bibnamefont {Hajek}}, \bibinfo {author} {\bibfnamefont {D.}~\bibnamefont {Salaj}}, \bibinfo {author} {\bibfnamefont {R.}~\bibnamefont {Legenstein}}, \ and\ \bibinfo {author} {\bibfnamefont {W.}~\bibnamefont {Maass}},\ }\bibfield  {title} {\enquote {\bibinfo {title} {Long short-term memory and learning-to-learn in networks of spiking neurons},}\ }in\ \href {\doibase https://anandsubramoney.com/files/pdfs/bellec_nips2018.pdf} {\emph {\bibinfo {booktitle} {NeurIPS}}},\ Vol.~\bibinfo {volume} {31}\ (\bibinfo {year} {2018})\BibitemShut {NoStop}%
\bibitem [{\citenamefont {Yadav}, \citenamefont {Sinha},\ and\ \citenamefont {Stender}(2025)}]{EvolveRC_Yadav2025}%
  \BibitemOpen
  \bibfield  {author} {\bibinfo {author} {\bibfnamefont {M.}~\bibnamefont {Yadav}}, \bibinfo {author} {\bibfnamefont {S.}~\bibnamefont {Sinha}}, \ and\ \bibinfo {author} {\bibfnamefont {M.}~\bibnamefont {Stender}},\ }\bibfield  {title} {\enquote {\bibinfo {title} {Performance-dependent network evolution for efficient information processing},}\ }\href {\doibase 10.1103/PhysRevE.111.014320} {\bibfield  {journal} {\bibinfo  {journal} {Physical Review E}\ }\textbf {\bibinfo {volume} {111}},\ \bibinfo {pages} {014320} (\bibinfo {year} {2025})}\BibitemShut {NoStop}%
\bibitem [{\citenamefont {Radhakrishnan}\ \emph {et~al.}(2025)\citenamefont {Radhakrishnan}, \citenamefont {Lindner}, \citenamefont {Miller}, \citenamefont {Sinha},\ and\ \citenamefont {Ditto}}]{Sinha2025}%
  \BibitemOpen
  \bibfield  {author} {\bibinfo {author} {\bibfnamefont {A.}~\bibnamefont {Radhakrishnan}}, \bibinfo {author} {\bibfnamefont {J.~F.}\ \bibnamefont {Lindner}}, \bibinfo {author} {\bibfnamefont {S.~T.}\ \bibnamefont {Miller}}, \bibinfo {author} {\bibfnamefont {S.}~\bibnamefont {Sinha}}, \ and\ \bibinfo {author} {\bibfnamefont {W.~L.}\ \bibnamefont {Ditto}},\ }\bibfield  {title} {\enquote {\bibinfo {title} {When less is more: evolving large neural networks from small ones},}\ }\href {\doibase https://arxiv.org/abs/2501.18012} {\bibfield  {journal} {\bibinfo  {journal} {arXiv preprint}\ }\textbf {\bibinfo {volume} {arXiv:2501.18012}} (\bibinfo {year} {2025}),\ https://arxiv.org/abs/2501.18012}\BibitemShut {NoStop}%
\bibitem [{\citenamefont {Rocks}\ \emph {et~al.}(2019)\citenamefont {Rocks}, \citenamefont {Pashine}, \citenamefont {Bischofberger}, \citenamefont {Goodrich}, \citenamefont {Liu},\ and\ \citenamefont {Nagel}}]{Rocks2019}%
  \BibitemOpen
  \bibfield  {author} {\bibinfo {author} {\bibfnamefont {J.~W.}\ \bibnamefont {Rocks}}, \bibinfo {author} {\bibfnamefont {N.}~\bibnamefont {Pashine}}, \bibinfo {author} {\bibfnamefont {I.}~\bibnamefont {Bischofberger}}, \bibinfo {author} {\bibfnamefont {C.~P.}\ \bibnamefont {Goodrich}}, \bibinfo {author} {\bibfnamefont {A.~J.}\ \bibnamefont {Liu}}, \ and\ \bibinfo {author} {\bibfnamefont {S.~R.}\ \bibnamefont {Nagel}},\ }\bibfield  {title} {\enquote {\bibinfo {title} {Designing allostery-inspired response in mechanical networks},}\ }\href {\doibase https://www.pnas.org/doi/10.1073/pnas.1612139114} {\bibfield  {journal} {\bibinfo  {journal} {Proceedings of the National Academy of Sciences (PNAS)}\ }\textbf {\bibinfo {volume} {116}},\ \bibinfo {pages} {2506--2511} (\bibinfo {year} {2019})}\BibitemShut {NoStop}%
\bibitem [{\citenamefont {Spielman}\ and\ \citenamefont {Teng}(2011)}]{Spielman2011}%
  \BibitemOpen
  \bibfield  {author} {\bibinfo {author} {\bibfnamefont {D.~A.}\ \bibnamefont {Spielman}}\ and\ \bibinfo {author} {\bibfnamefont {S.-H.}\ \bibnamefont {Teng}},\ }\bibfield  {title} {\enquote {\bibinfo {title} {Spectral sparsification of graphs},}\ }\href {\doibase https://epubs.siam.org/doi/10.1137/08074489X} {\bibfield  {journal} {\bibinfo  {journal} {SIAM Journal on Computing}\ }\textbf {\bibinfo {volume} {40}},\ \bibinfo {pages} {981--1025} (\bibinfo {year} {2011})}\BibitemShut {NoStop}%
\bibitem [{\citenamefont {Han}\ \emph {et~al.}(2015)\citenamefont {Han}, \citenamefont {Pool}, \citenamefont {Tran},\ and\ \citenamefont {Dally}}]{Han2015}%
  \BibitemOpen
  \bibfield  {author} {\bibinfo {author} {\bibfnamefont {S.}~\bibnamefont {Han}}, \bibinfo {author} {\bibfnamefont {J.}~\bibnamefont {Pool}}, \bibinfo {author} {\bibfnamefont {J.}~\bibnamefont {Tran}}, \ and\ \bibinfo {author} {\bibfnamefont {W.}~\bibnamefont {Dally}},\ }\bibfield  {title} {\enquote {\bibinfo {title} {Learning both weights and connections for efficient neural networks},}\ }in\ \href {\doibase 10.5555/3044805.3044999} {\emph {\bibinfo {booktitle} {Advances in Neural Information Processing Systems (NeurIPS)}}},\ Vol.~\bibinfo {volume} {28}\ (\bibinfo {year} {2015})\BibitemShut {NoStop}%
\bibitem [{\citenamefont {Molchanov}\ \emph {et~al.}(2017)\citenamefont {Molchanov}, \citenamefont {Tyree}, \citenamefont {Karras}, \citenamefont {Aila},\ and\ \citenamefont {Kautz}}]{Molchanov2017}%
  \BibitemOpen
  \bibfield  {author} {\bibinfo {author} {\bibfnamefont {P.}~\bibnamefont {Molchanov}}, \bibinfo {author} {\bibfnamefont {S.}~\bibnamefont {Tyree}}, \bibinfo {author} {\bibfnamefont {T.}~\bibnamefont {Karras}}, \bibinfo {author} {\bibfnamefont {T.}~\bibnamefont {Aila}}, \ and\ \bibinfo {author} {\bibfnamefont {J.}~\bibnamefont {Kautz}},\ }\bibfield  {title} {\enquote {\bibinfo {title} {Pruning convolutional neural networks for resource-efficient inference},}\ }in\ \href {\doibase 10.1109/CVPR.2017.567} {\emph {\bibinfo {booktitle} {International Conference on Learning Representations (ICLR)}}}\ (\bibinfo {year} {2017})\BibitemShut {NoStop}%
\bibitem [{\citenamefont {Mead}(1989)}]{Mead1990}%
  \BibitemOpen
  \bibfield  {author} {\bibinfo {author} {\bibfnamefont {C.}~\bibnamefont {Mead}},\ }\href@noop {} {\emph {\bibinfo {title} {Analog VLSI and Neural Systems}}}\ (\bibinfo  {publisher} {Addison-Wesley},\ \bibinfo {year} {1989})\BibitemShut {NoStop}%
\bibitem [{\citenamefont {Hussain}\ and\ \citenamefont {Li}(2023)}]{Hussain2023}%
  \BibitemOpen
  \bibfield  {author} {\bibinfo {author} {\bibfnamefont {I.}~\bibnamefont {Hussain}}\ and\ \bibinfo {author} {\bibfnamefont {T.}~\bibnamefont {Li}},\ }\bibfield  {title} {\enquote {\bibinfo {title} {An optimization-based algorithm for obtaining an effective sparse network},}\ }\href {\doibase 10.1063/5.0131692} {\bibfield  {journal} {\bibinfo  {journal} {Chaos}\ }\textbf {\bibinfo {volume} {33}},\ \bibinfo {pages} {033103} (\bibinfo {year} {2023})}\BibitemShut {NoStop}%
\bibitem [{\citenamefont {Yan}\ and\ \citenamefont {Wang}(2023)}]{Yan2023}%
  \BibitemOpen
  \bibfield  {author} {\bibinfo {author} {\bibfnamefont {Z.}~\bibnamefont {Yan}}\ and\ \bibinfo {author} {\bibfnamefont {H.}~\bibnamefont {Wang}},\ }\bibfield  {title} {\enquote {\bibinfo {title} {A perturbation-based approach to identifying unnecessary nodes in neural networks},}\ }\href {\doibase 10.1063/5.0144713} {\bibfield  {journal} {\bibinfo  {journal} {Chaos}\ }\textbf {\bibinfo {volume} {33}},\ \bibinfo {pages} {063119} (\bibinfo {year} {2023})}\BibitemShut {NoStop}%
\bibitem [{\citenamefont {Jiang}\ and\ \citenamefont {Zhao}(2024)}]{Jiang2024}%
  \BibitemOpen
  \bibfield  {author} {\bibinfo {author} {\bibfnamefont {L.}~\bibnamefont {Jiang}}\ and\ \bibinfo {author} {\bibfnamefont {R.}~\bibnamefont {Zhao}},\ }\bibfield  {title} {\enquote {\bibinfo {title} {Keeping some adaptivity in the network after pruning},}\ }\href {\doibase 10.1063/5.0211692} {\bibfield  {journal} {\bibinfo  {journal} {Chaos}\ }\textbf {\bibinfo {volume} {34}},\ \bibinfo {pages} {012345} (\bibinfo {year} {2024})}\BibitemShut {NoStop}%
\bibitem [{\citenamefont {Hasani}, \citenamefont {Lechner},\ and\ \citenamefont {Grosu}(2023)}]{Hasani2023}%
  \BibitemOpen
  \bibfield  {author} {\bibinfo {author} {\bibfnamefont {R.}~\bibnamefont {Hasani}}, \bibinfo {author} {\bibfnamefont {M.}~\bibnamefont {Lechner}}, \ and\ \bibinfo {author} {\bibfnamefont {R.}~\bibnamefont {Grosu}},\ }\href {https://www.liquid.ai/research/automated-architecture-synthesis-via-targeted-evolution} {\enquote {\bibinfo {title} {Automated architecture synthesis via targeted evolution},}\ }\bibinfo {howpublished} {Liquid AI Research} (\bibinfo {year} {2023})\BibitemShut {NoStop}%
\bibitem [{\citenamefont {Jaeger}(2001)}]{Jaeger2001}%
  \BibitemOpen
  \bibfield  {author} {\bibinfo {author} {\bibfnamefont {H.}~\bibnamefont {Jaeger}},\ }\href {https://www.ai.mit.edu/projects/dynamical-systems/jaeger2001echo.pdf} {\enquote {\bibinfo {title} {The `echo state' approach to analyzing and training recurrent neural networks},}\ }\bibinfo {type} {Tech. Rep.}\ \bibinfo {number} {GMD Technical Report 148}\ (\bibinfo  {institution} {German National Research Center for Information Technology},\ \bibinfo {year} {2001})\BibitemShut {NoStop}%
\bibitem [{\citenamefont {Verstraeten}\ \emph {et~al.}(2007)\citenamefont {Verstraeten}, \citenamefont {Schrauwen}, \citenamefont {D'Haene},\ and\ \citenamefont {Stroobandt}}]{Verstraeten2007}%
  \BibitemOpen
  \bibfield  {author} {\bibinfo {author} {\bibfnamefont {D.}~\bibnamefont {Verstraeten}}, \bibinfo {author} {\bibfnamefont {B.}~\bibnamefont {Schrauwen}}, \bibinfo {author} {\bibfnamefont {M.}~\bibnamefont {D'Haene}}, \ and\ \bibinfo {author} {\bibfnamefont {D.}~\bibnamefont {Stroobandt}},\ }\bibfield  {title} {\enquote {\bibinfo {title} {An experimental unification of reservoir computing methods},}\ }\href {\doibase https://doi.org/10.1016/j.neunet.2007.04.003} {\bibfield  {journal} {\bibinfo  {journal} {Neural Networks}\ }\textbf {\bibinfo {volume} {20}},\ \bibinfo {pages} {391--403} (\bibinfo {year} {2007})}\BibitemShut {NoStop}%
\bibitem [{\citenamefont {Yildiz}, \citenamefont {Jaeger},\ and\ \citenamefont {Kiebel}(2012)}]{Yildiz2012}%
  \BibitemOpen
  \bibfield  {author} {\bibinfo {author} {\bibfnamefont {I.~B.}\ \bibnamefont {Yildiz}}, \bibinfo {author} {\bibfnamefont {H.}~\bibnamefont {Jaeger}}, \ and\ \bibinfo {author} {\bibfnamefont {S.~J.}\ \bibnamefont {Kiebel}},\ }\bibfield  {title} {\enquote {\bibinfo {title} {Re-visiting the echo state property},}\ }\href {\doibase https://doi.org/10.1016/j.neunet.2012.07.005} {\bibfield  {journal} {\bibinfo  {journal} {Neural Networks}\ }\textbf {\bibinfo {volume} {35}},\ \bibinfo {pages} {1--9} (\bibinfo {year} {2012})}\BibitemShut {NoStop}%
\bibitem [{\citenamefont {Gallicchio}, \citenamefont {Micheli},\ and\ \citenamefont {Pedrelli}(2017)}]{Gallicchio2017}%
  \BibitemOpen
  \bibfield  {author} {\bibinfo {author} {\bibfnamefont {C.}~\bibnamefont {Gallicchio}}, \bibinfo {author} {\bibfnamefont {A.}~\bibnamefont {Micheli}}, \ and\ \bibinfo {author} {\bibfnamefont {L.}~\bibnamefont {Pedrelli}},\ }\bibfield  {title} {\enquote {\bibinfo {title} {Deep reservoir computing: A critical experimental analysis},}\ }\href {\doibase https://doi.org/10.1016/j.neucom.2016.12.089} {\bibfield  {journal} {\bibinfo  {journal} {Neurocomputing}\ }\textbf {\bibinfo {volume} {268}},\ \bibinfo {pages} {87--99} (\bibinfo {year} {2017})}\BibitemShut {NoStop}%
\bibitem [{\citenamefont {Lukosevicius}, \citenamefont {Jaeger},\ and\ \citenamefont {Schrauwen}(2012)}]{Lukosevicius2012}%
  \BibitemOpen
  \bibfield  {author} {\bibinfo {author} {\bibfnamefont {M.}~\bibnamefont {Lukosevicius}}, \bibinfo {author} {\bibfnamefont {H.}~\bibnamefont {Jaeger}}, \ and\ \bibinfo {author} {\bibfnamefont {B.}~\bibnamefont {Schrauwen}},\ }\bibfield  {title} {\enquote {\bibinfo {title} {Reservoir computing trends},}\ }in\ \href {\doibase 10.1007/978-3-319-18164-6_4} {\emph {\bibinfo {booktitle} {Lecture Notes in Computer Science}}},\ Vol.\ \bibinfo {volume} {718}\ (\bibinfo  {publisher} {Springer},\ \bibinfo {year} {2012})\ pp.\ \bibinfo {pages} {384--397}\BibitemShut {NoStop}%
\bibitem [{\citenamefont {Schrauwen}, \citenamefont {Verstraeten},\ and\ \citenamefont {Van~Campenhout}(2009)}]{Schrauwen2009}%
  \BibitemOpen
  \bibfield  {author} {\bibinfo {author} {\bibfnamefont {B.}~\bibnamefont {Schrauwen}}, \bibinfo {author} {\bibfnamefont {D.}~\bibnamefont {Verstraeten}}, \ and\ \bibinfo {author} {\bibfnamefont {J.}~\bibnamefont {Van~Campenhout}},\ }\bibfield  {title} {\enquote {\bibinfo {title} {Pruning reservoir computing networks},}\ }\href {\doibase 10.1016/j.neucom.2008.10.013} {\bibfield  {journal} {\bibinfo  {journal} {Neurocomputing}\ }\textbf {\bibinfo {volume} {72}},\ \bibinfo {pages} {1534--1546} (\bibinfo {year} {2009})}\BibitemShut {NoStop}%
\bibitem [{\citenamefont {Can}\ and\ \citenamefont {Gauthier}(2024)}]{Can2024}%
  \BibitemOpen
  \bibfield  {author} {\bibinfo {author} {\bibfnamefont {Y.}~\bibnamefont {Can}}\ and\ \bibinfo {author} {\bibfnamefont {D.}~\bibnamefont {Gauthier}},\ }\bibfield  {title} {\enquote {\bibinfo {title} {Structural evolution in reservoir computing networks},}\ }\href {\doibase 10.1038/s44172-024-00227-y} {\bibfield  {journal} {\bibinfo  {journal} {Nature Machine Intelligence}\ }\textbf {\bibinfo {volume} {6}},\ \bibinfo {pages} {202--215} (\bibinfo {year} {2024})}\BibitemShut {NoStop}%
\bibitem [{\citenamefont {Schiller}, \citenamefont {Hülser},\ and\ \citenamefont {Lindner}(2020)}]{Schiller2020}%
  \BibitemOpen
  \bibfield  {author} {\bibinfo {author} {\bibfnamefont {L.}~\bibnamefont {Schiller}}, \bibinfo {author} {\bibfnamefont {T.}~\bibnamefont {Hülser}}, \ and\ \bibinfo {author} {\bibfnamefont {B.}~\bibnamefont {Lindner}},\ }\bibfield  {title} {\enquote {\bibinfo {title} {The effect of pruning on the performance of echo state networks},}\ }\href {\doibase 10.1063/5.0006869} {\bibfield  {journal} {\bibinfo  {journal} {Chaos}\ }\textbf {\bibinfo {volume} {30}},\ \bibinfo {pages} {045678} (\bibinfo {year} {2020})}\BibitemShut {NoStop}%
\bibitem [{Note1()}]{Note1}%
  \BibitemOpen
  \bibinfo {note} {\unhbox \voidb@x \begingroup \begingroup \let \relax \relax \relax \endgroup \protect \Url {https://github.com/Cyber-Physical-Systems-in-Mech-Eng/pyReCo}}\BibitemShut {NoStop}%
\bibitem [{\citenamefont {Atiya}\ and\ \citenamefont {Parlos}(2000)}]{Atiya2000}%
  \BibitemOpen
  \bibfield  {author} {\bibinfo {author} {\bibfnamefont {A.}~\bibnamefont {Atiya}}\ and\ \bibinfo {author} {\bibfnamefont {A.}~\bibnamefont {Parlos}},\ }\bibfield  {title} {\enquote {\bibinfo {title} {New results on recurrent network training: unifying the algorithms and accelerating convergence},}\ }\href {\doibase https://doi.org/10.1109/72.846741} {\bibfield  {journal} {\bibinfo  {journal} {IEEE Transactions on Neural Networks}\ }\textbf {\bibinfo {volume} {11}},\ \bibinfo {pages} {697} (\bibinfo {year} {2000})}\BibitemShut {NoStop}%
\bibitem [{\citenamefont {Maass}, \citenamefont {Natschlaeger},\ and\ \citenamefont {Markram}(2002)}]{Maass2002}%
  \BibitemOpen
  \bibfield  {author} {\bibinfo {author} {\bibfnamefont {W.}~\bibnamefont {Maass}}, \bibinfo {author} {\bibfnamefont {T.}~\bibnamefont {Natschlaeger}}, \ and\ \bibinfo {author} {\bibfnamefont {H.}~\bibnamefont {Markram}},\ }\bibfield  {title} {\enquote {\bibinfo {title} {Real-time computing without stable states: A new framework for neural computation based on perturbations},}\ }\href {\doibase 10.1162/089976602760407955} {\bibfield  {journal} {\bibinfo  {journal} {Neural Computation}\ }\textbf {\bibinfo {volume} {14}},\ \bibinfo {pages} {2531} (\bibinfo {year} {2002})}\BibitemShut {NoStop}%
\bibitem [{\citenamefont {Pathak}\ \emph {et~al.}(2018)\citenamefont {Pathak}, \citenamefont {Hunt}, \citenamefont {Girvan}, \citenamefont {Lu},\ and\ \citenamefont {Ott}}]{Pathak2018}%
  \BibitemOpen
  \bibfield  {author} {\bibinfo {author} {\bibfnamefont {J.}~\bibnamefont {Pathak}}, \bibinfo {author} {\bibfnamefont {B.~R.}\ \bibnamefont {Hunt}}, \bibinfo {author} {\bibfnamefont {M.}~\bibnamefont {Girvan}}, \bibinfo {author} {\bibfnamefont {Z.}~\bibnamefont {Lu}}, \ and\ \bibinfo {author} {\bibfnamefont {E.}~\bibnamefont {Ott}},\ }\bibfield  {title} {\enquote {\bibinfo {title} {Model-free prediction of large spatiotemporally chaotic systems from data},}\ }\href {\doibase 10.1126/science.aar4193} {\bibfield  {journal} {\bibinfo  {journal} {Science}\ }\textbf {\bibinfo {volume} {360}},\ \bibinfo {pages} {786--790} (\bibinfo {year} {2018})}\BibitemShut {NoStop}%
\bibitem [{\citenamefont {Gauthier}(2021)}]{Gauthier2021}%
  \BibitemOpen
  \bibfield  {author} {\bibinfo {author} {\bibfnamefont {J.}~\bibnamefont {Gauthier}},\ }\bibfield  {title} {\enquote {\bibinfo {title} {Emergent properties of machine learning in dynamic systems},}\ }\href {\doibase https://doi.org/10.1038/s41467-021-21109-w} {\bibfield  {journal} {\bibinfo  {journal} {Nature Communications}\ }\textbf {\bibinfo {volume} {12}},\ \bibinfo {pages} {5439} (\bibinfo {year} {2021})}\BibitemShut {NoStop}%
\bibitem [{\citenamefont {Huttenlocher}(1979)}]{Huttenlocher1979}%
  \BibitemOpen
  \bibfield  {author} {\bibinfo {author} {\bibfnamefont {J.}~\bibnamefont {Huttenlocher}},\ }\bibfield  {title} {\enquote {\bibinfo {title} {The nature of cognitive development: A psychological perspective},}\ }\href {\doibase https://doi.org/10.1037/0033-295X.86.4.331} {\bibfield  {journal} {\bibinfo  {journal} {Psychological Review}\ }\textbf {\bibinfo {volume} {86}},\ \bibinfo {pages} {331--347} (\bibinfo {year} {1979})}\BibitemShut {NoStop}%
\bibitem [{\citenamefont {Chechik}\ \emph {et~al.}(1998)\citenamefont {Chechik} \emph {et~al.}}]{Chechik1998}%
  \BibitemOpen
  \bibfield  {author} {\bibinfo {author} {\bibfnamefont {G.}~\bibnamefont {Chechik}} \emph {et~al.},\ }\bibfield  {title} {\enquote {\bibinfo {title} {An integrative theory of memory and perception},}\ }\href {\doibase https://doi.org/10.1162/089892998563798} {\bibfield  {journal} {\bibinfo  {journal} {Journal of Cognitive Neuroscience}\ }\textbf {\bibinfo {volume} {10}},\ \bibinfo {pages} {410--430} (\bibinfo {year} {1998})}\BibitemShut {NoStop}%
\bibitem [{\citenamefont {Alon}(2006)}]{Alon2006}%
  \BibitemOpen
  \bibfield  {author} {\bibinfo {author} {\bibfnamefont {U.}~\bibnamefont {Alon}},\ }\bibfield  {title} {\enquote {\bibinfo {title} {Network motifs: Theory and experimental approaches},}\ }\href {\doibase https://doi.org/10.1038/nrg1815} {\bibfield  {journal} {\bibinfo  {journal} {Nature Reviews Genetics}\ }\textbf {\bibinfo {volume} {7}},\ \bibinfo {pages} {4--16} (\bibinfo {year} {2006})}\BibitemShut {NoStop}%
\bibitem [{\citenamefont {Rodan}\ and\ \citenamefont {Tino}(2011)}]{Rodan2011}%
  \BibitemOpen
  \bibfield  {author} {\bibinfo {author} {\bibfnamefont {A.}~\bibnamefont {Rodan}}\ and\ \bibinfo {author} {\bibfnamefont {P.}~\bibnamefont {Tino}},\ }\bibfield  {title} {\enquote {\bibinfo {title} {Minimum complexity echo state networks},}\ }\href {\doibase https://doi.org/10.1109/TNN.2010.2105762} {\bibfield  {journal} {\bibinfo  {journal} {IEEE Transactions on Neural Networks}\ }\textbf {\bibinfo {volume} {22}},\ \bibinfo {pages} {131--144} (\bibinfo {year} {2011})}\BibitemShut {NoStop}%
\bibitem [{\citenamefont {Schrauwen}\ \emph {et~al.}(2008)\citenamefont {Schrauwen} \emph {et~al.}}]{Schrauwen2008}%
  \BibitemOpen
  \bibfield  {author} {\bibinfo {author} {\bibfnamefont {B.}~\bibnamefont {Schrauwen}} \emph {et~al.},\ }\bibfield  {title} {\enquote {\bibinfo {title} {The echo state network: A powerful tool for the modeling of dynamic systems},}\ }\href {\doibase https://doi.org/10.1016/j.neunet.2008.03.008} {\bibfield  {journal} {\bibinfo  {journal} {Neural Networks}\ }\textbf {\bibinfo {volume} {21}},\ \bibinfo {pages} {1212--1221} (\bibinfo {year} {2008})}\BibitemShut {NoStop}%
\bibitem [{\citenamefont {LeCun}\ \emph {et~al.}(1990)\citenamefont {LeCun}, \citenamefont {Boser}, \citenamefont {Denker}, \citenamefont {Henderson}, \citenamefont {Howard}, \citenamefont {Hubbard},\ and\ \citenamefont {Jackel}}]{LeCun1990}%
  \BibitemOpen
  \bibfield  {author} {\bibinfo {author} {\bibfnamefont {Y.}~\bibnamefont {LeCun}}, \bibinfo {author} {\bibfnamefont {B.}~\bibnamefont {Boser}}, \bibinfo {author} {\bibfnamefont {J.~S.}\ \bibnamefont {Denker}}, \bibinfo {author} {\bibfnamefont {D.}~\bibnamefont {Henderson}}, \bibinfo {author} {\bibfnamefont {R.~E.}\ \bibnamefont {Howard}}, \bibinfo {author} {\bibfnamefont {W.}~\bibnamefont {Hubbard}}, \ and\ \bibinfo {author} {\bibfnamefont {L.~D.}\ \bibnamefont {Jackel}},\ }\bibfield  {title} {\enquote {\bibinfo {title} {Handwritten digit recognition with a back-propagation network},}\ }in\ \href {\doibase 10.1109/ICASSP.1990.115055} {\emph {\bibinfo {booktitle} {IEEE Proceedings of the International Conference on Acoustics, Speech, and Signal Processing}}},\ Vol.~\bibinfo {volume} {1}\ (\bibinfo {year} {1990})\ pp.\ \bibinfo {pages} {541--544}\BibitemShut {NoStop}%
\bibitem [{\citenamefont {LeCun}\ \emph {et~al.}(1998)\citenamefont {LeCun}, \citenamefont {Bottou}, \citenamefont {Bengio},\ and\ \citenamefont {Haffner}}]{LeCun1998}%
  \BibitemOpen
  \bibfield  {author} {\bibinfo {author} {\bibfnamefont {Y.}~\bibnamefont {LeCun}}, \bibinfo {author} {\bibfnamefont {L.}~\bibnamefont {Bottou}}, \bibinfo {author} {\bibfnamefont {Y.}~\bibnamefont {Bengio}}, \ and\ \bibinfo {author} {\bibfnamefont {P.}~\bibnamefont {Haffner}},\ }\bibfield  {title} {\enquote {\bibinfo {title} {Gradient-based learning applied to document recognition},}\ }\href {\doibase https://ieeexplore.ieee.org/document/726791} {\bibfield  {journal} {\bibinfo  {journal} {Proceedings of the IEEE}\ }\textbf {\bibinfo {volume} {86}},\ \bibinfo {pages} {2278--2324} (\bibinfo {year} {1998})}\BibitemShut {NoStop}%
\bibitem [{\citenamefont {Rumelhart}, \citenamefont {Hinton},\ and\ \citenamefont {Williams}(1986)}]{Rumelhart1986}%
  \BibitemOpen
  \bibfield  {author} {\bibinfo {author} {\bibfnamefont {D.~E.}\ \bibnamefont {Rumelhart}}, \bibinfo {author} {\bibfnamefont {G.~E.}\ \bibnamefont {Hinton}}, \ and\ \bibinfo {author} {\bibfnamefont {R.~J.}\ \bibnamefont {Williams}},\ }\bibfield  {title} {\enquote {\bibinfo {title} {Learning representations by back-propagating errors},}\ }\href {\doibase https://www.nature.com/articles/323533a0} {\bibfield  {journal} {\bibinfo  {journal} {Nature}\ }\textbf {\bibinfo {volume} {323}},\ \bibinfo {pages} {533--536} (\bibinfo {year} {1986})}\BibitemShut {NoStop}%
\bibitem [{\citenamefont {Lukoševičius}(2012)}]{Lukosevicius2012Guide}%
  \BibitemOpen
  \bibfield  {author} {\bibinfo {author} {\bibfnamefont {M.}~\bibnamefont {Lukoševičius}},\ }\bibfield  {title} {\enquote {\bibinfo {title} {A practical guide to applying echo state networks},}\ }in\ \href {\doibase https://doi.org/10.1007/978-3-642-35289-8_36} {\emph {\bibinfo {booktitle} {Neural Networks: Tricks of the Trade}}},\ \bibinfo {series} {Lecture Notes in Computer Science (LNCS)}, Vol.\ \bibinfo {volume} {7700}\ (\bibinfo  {publisher} {Springer, Heidelberg},\ \bibinfo {year} {2012})\ \bibinfo {edition} {2nd}\ ed.,\ pp.\ \bibinfo {pages} {659--686}\BibitemShut {NoStop}%
\end{thebibliography}
%merlin.mbs aipnum4-1.bst 2010-07-25 4.21a (PWD, AO, DPC) hacked
%Control: key (0)
%Control: author (8) initials jnrlst
%Control: editor formatted (1) identically to author
%Control: production of article title (0) allowed
%Control: page (1) range
%Control: year (1) truncated
%Control: production of eprint (0) enabled
%

\end{document}